# Systematic Review of Techniques in Brain Image Synthesis using Deep Learning

Shubham Singh[1], Ammar Ranapurwala[1], Mrunal Bewoor[1], Sheetal Patil[1], Satyam Rai[1]

1 Bharati Vidyapeeth (Deemed to be University) College of Engineering, Pune

**Abstract** — This review paper delves into the present state of medical imaging, with a specific focus on the use of deep learning techniques for brain image synthesis. The need for medical image synthesis to improve diagnostic accuracy and decrease invasiveness in medical procedures is emphasized, along with the role of deep learning in enabling these advancements. The paper examines various methods and techniques for brain image synthesis, including 2D to 3D constructions, MRI synthesis, and the use of transformers. It also addresses limitations and challenges faced in these methods, such as obtaining well-curated training data and addressing brain ultrasound issues. The review concludes by exploring the future potential of this field and the opportunities for further advancements in medical imaging using deep learning techniques. The significance of transformers and their potential to revolutionize the medical imaging field is highlighted. Additionally, the paper discusses the potential solutions to the shortcomings and limitations faced in this field. The review provides researchers with an updated reference on the present state of the field and aims to inspire further research and bridge the gap between the present state of medical imaging and the future possibilities offered by deep learning techniques.

## I. INTRODUCTION

The review discusses significance of medical imaging in the diagnosis and treatment planning of various diseases and the various imaging modalities utilized. Other topics discussed includes tools and architectures developed in deep learning around medical applications.

Medical imaging process can be dissected into several steps involving capturing the image, processing the image, synthesizing the image if necessary, and finally using an algorithm to run a diagnosis on it. The review breaks down these processes and talks about innovations in deep learning that are enhancing these processes. Additionally, the review highlights the challenges associated with these processes such as cost and associated risks. Conventional synthesis approaches utilize nonlinear models, such as dictionary learning and random forest, to process handcrafted medical image features selected by experts. However, these manual features have a limited ability to represent the complex information present in medical images, thus affecting synthesis performance.

Deep learning-based methods have been successful in addressing these limitations by automatically learning features with sufficient descriptive power through the training of mapping models. These advanced deep-learning models have improved the quality of medical image synthesis and decreased the associated costs. The literature review will examine recent research in this field and discuss the potential for deep learning-based medical image synthesis to improve diagnostic accuracy and reduce invasiveness in medical procedures.

In recent years, medical imaging field has seen significant advancements with the increasing use of deep learning techniques enhancing diagnostic accuracy and reducing the invasiveness of medical procedures. The studies reviewed in this section provide an overview of recent developments in deep learning applications for medical image analysis.

In context to the brain, its applications include brain tumour segmentation, lesion identification, and radiation therapy planning (Sahiner et al.,2019). Deep neural networks have come to outperform human capabilities in computer vision tasks such as disease diagnosis and are being explored to deliver precision medicine (Kim et al., 2019). Several tools, architectures, and algorithms have been developed to aid research in the field. "A review of the application of deep learning in medical image classification and segmentation" talks about them in detail. One of the most popular structures for image classification is the convolutional neural network (CNN). AlexNet, a CNN-based deep learning model, was proposed in 2012 which popularized CNNs. Other popular CNN structures include the network in the network (NIN), GoogLeNet, VGGNet, SegNet, U-Net, and ResNet. Some popular deep-learning frameworks used in the field of medical imaging include Caffe, TensorFlow, and PyTorch. Caffe is known for its high performance, seamless switching between CPU and GPU modes, and support for multiple platforms. TensorFlow is a library which is open source, widely used and provides powerful visualization capabilities and support for heterogeneous distributed computing. Pytorch is specifically targeted at GPU-accelerated deep neural network programming, and it has a dynamic calculation graph that can be changed in real-time. These advancements have greatly aided the field of deep-learning image research in medical imaging (Cai et. al., 2020).

## II. EXISTING WORKS

*A. Image Processing:*

There are a lot of state-of-the-art deep learning architectures for segmentation and classification of medical images which uses techniques to extract information from images and present it in an efficient and effective form, which can assist doctors in diagnosing and predicting the risk of diseases more accurately.

But there are also challenges and research issues related to the use of deep learning in medical imaging including the availability of big data, recent deep learning algorithms modelled on the human brain, and processing power (Razzak et al., 2017). To overcome the issue of big data unavailability the field of supervised deep Learning is required to shift from supervised to unsupervised or semi-supervised methods as deep learning applications rely on extremely large datasets, and the availability of such data is not always possible.

The study on the topic by Andreasen talks about the basic issues with the analysis of structural and functional imaging data in neuroimaging studies. Data transmission, boundary detection, volume estimates, 3-D reconstruction and presentation, surface and volume rendering, shape analysis, and picture overlay are some of these issues.

The Study asserts that these problems require application of different methods of image analysis, implemented on a set of software programs, to conduct neuroimaging research using magnetic resonance, single-photon emission computed tomography (Greenwood et. al., 2012). It also describes a group of software programs called BRAINS, designed to provide a comprehensive solution for these problems (Andreasen et al., 1992).

Noise reduction in medical images is important because noise can interfere with the interpretation of the image and lead to inaccurate diagnoses or treatment decisions. Noise may result from a variety of factors such as electronic noise from image sensors, quantization errors, and compression artifacts. Noise can make it difficult to distinguish between important structures or features in an image, making it harder to make a diagnosis or treatment decision. Additionally, noise can reduce the visibility of small or subtle structures or features in an image, which can be critical for accurate diagnoses or treatment decisions. Furthermore, noise can also increase the risk of false positives or false negatives, which can have serious consequences for patients. Therefore, it is crucial to reduce noise in medical images to ensure accurate and reliable diagnoses or treatment decisions. The introduction of methods like fuzzy filters has provided a simple and efficient option for noise reduction of medical images (Van De Ville et al., 2003).

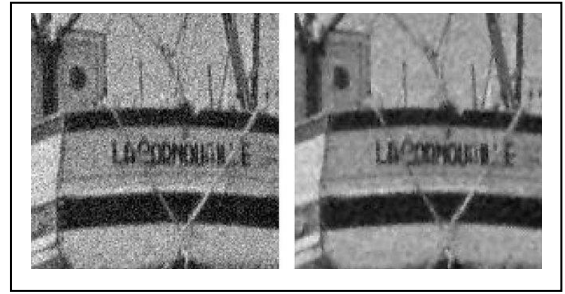

Fig 1. Noise reduction using fuzzy filters (Van de Ville et. al. 2003)

Qualitative and quantitative volumetric analysis of brain images can be carried out for the development, implementation, and validation of an image-processing system. The system allows for the visualization and quantitation of global and regional brain volumes and uses techniques such as automated adaptive Bayesian segmentation, normalization, and tessellation of segmented brain images into the Talairach space, and a hybrid method combining a region-of-interest approach and voxel-based analysis. The goal of this image-processing system is to measure the magnitude, rate, and regional pattern of longitudinal changes in the brain, and it could be useful in understanding and detecting presence of abnormalities in brain structure due to aging or disease (Goldszal et al., 1998). The below figure illustrates a system overview of the entire model.

The development and implementation of an automated image processing and Quality Control pipeline for the brain images obtained from various modalities using imaging protocol and processing pipelines (Fidel Alfaro-Almagro et al., 2016). The paper proposes a lot of great pipelines designed to help convert the raw data from the imaged subjects into useful summary information and is available for use by other researchers.

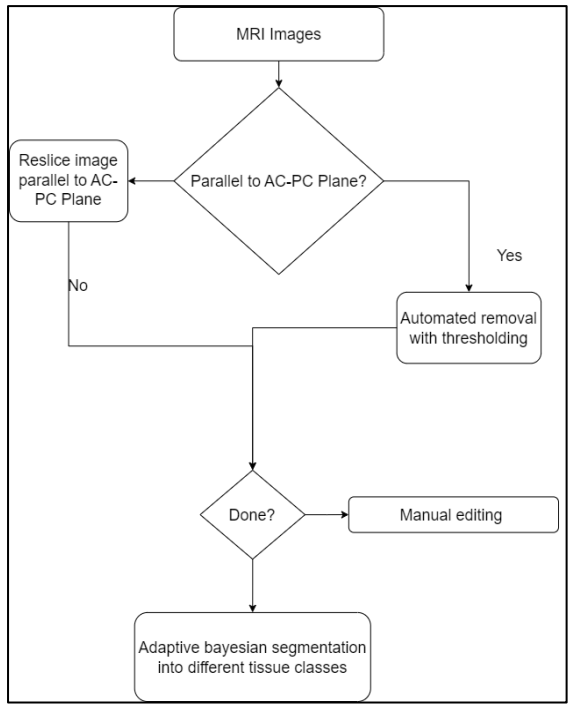

Fig 2. The image processing system

*B. Addressing problems with ultrasound imaging*

Brain ultrasound scans are difficult to capture with accuracy simply because the skull blocks some of the sound signals, thus adding in noise. The image processing techniques discussed above are a crucial step to removing that noise.

There are several other modalities that try to tackle the problem in an alternative way.

Transcranial Doppler (TCD) ultrasonography has been used as a non-invasive method for measuring blood flow and cerebrovascular hemodynamic within the basal arteries of the brain. TCD has limitations such as its operator dependency and the inadequate acoustic windows prevalent in certain populations, which can hinder its more widespread use. It is concluded that TCD is an essential tool that can be used in the diagnosis and management of various cerebrovascular disorders and in research settings. (Purkayastha and Sorond et al., 2013).

Researchers reviewed the efforts to circumvent the blood-brain barrier through the design of new drugs and the development of more sophisticated delivery methods which additionally highlighted the recent advances in the development of non-invasive, targeted drug delivery by MRI-guided ultrasound-induced BBB disruption. However, it also acknowledged the limitations of focused ultrasound, such as the need for an acoustic window made by a craniotomy and the trade-off between frequency and focal spot size (Vykhodtseva, McDannold, and Hynynen et al., 2008).

*C. 2D to 3D constructions:*

After discussions into image acquisition and image processing, recent development, has been to create 3D images of medical images, this helps us understand the images better and make better decisions. This is particularly difficult to do for a regular patient due to the need for several slowly captured images in a linear axial motion and then, combining them into a 3D scan. There are papers highlighting the problem and trying to solve them.

2D to 3D image reconstruction is a technique used in medical imaging to convert a series of 2D images into a single 3D image. The techniques used include multi-planar reformatting, Volume rendering, Maximum intensity projection, Surface rendering, Shape-from-shading, Cone-beam CT, Model-based algorithms, and Deep Learning-based methods. 2D to 3D reconstruction can be used in various medical imaging modalities such as CT, MRI, and ultrasound. The benefits include improved diagnostic accuracy and treatment planning, but it also has some limitations such as increased computation time, storage requirements, and reliance on the quality of the 2D images.

Heimann et. al. proposed several new ideas in the field like the use of Statistical shape models as a robust tool for the segmentation of medical images. It considers the methods required to construct and use these 3D SSMs while concentrating on landmark-based shape representations and thoroughly examining the most popular variants of Active Shape and Active Appearance models. It also describes several alternative approaches to statistical shape modelling. The paper discusses shape representation and correspondence, model construction, local appearance models, and search algorithms, and provides an overview of the current state of the art in the field.

It highlights that establishing dense point correspondences between all shapes of the training set is the most challenging part of 3D model construction and one of the major factors influencing model quality. It also discusses that manual landmarking is getting increasingly unpopular due to the tedious and time-consuming expert work required and the lack of reproducibility of the results. A study on the topic by Krucker describes a new technique called 3D spatial compounding of ultrasound images using image-based nonrigid registration. The technique is used to overcome the limitations of resolution during the compounding of ultrasound images. The method uses volumetric ultrasound data acquired by scanning a linear matrix array probe in the elevational direction in a focal lesion phantom and in a breast in vivo. The study shows that the technique is successful in enabling high spatial resolution in 3D spatial compounding of ultrasound images. (Krucker and Charles et.al, 2000).

*D. Medical image Synthesis:*

A Deep learning-based approach can be taken for MRI synthesis from brain computed tomography (CT) images for magnetic resonance (MR)-guided radiotherapy. A method can be employed for synthesizing brain MRI images from corresponding planning CT (pCT) images using deep learning methods. A mixture of different deep learning models were applied to implement this task, including CycleGAN, Pix2Pix model, and U-Net. It evaluated these methods using several metrics, including mean absolute error (MAE), mean squared error (MSE), structural similarity index (SSIM), and peak signal-to-noise ratio (PSNR). Overall, the authors conclude that the proposed method has the potential to improve patient positioning in radiotherapy by generating synthetic brain MRI images from CT (Li et al., 2020).

A novel approach can be taken for cross-modality MRI synthesis using a generative adversarial network (GAN) called EP_IMF-GAN. The GAN is designed to preserve edge information, which is critical for providing clinical information, by incorporating an auxiliary task of generating the corresponding edge image of the target modality. This is achieved by enforcing the primary task of generating the target modality and the auxiliary task of generating the edge image to share the same encoder, thus allowing the primary task to be supplemented with the domain-specific information learned from the auxiliary task. Additionally, an iterative multi-scale fusion module is embedded in the primary decoder to further improve the quality of the synthesized target modality. Results from experiments on the BRATS dataset show that the proposed EP_IMF-GAN method outperforms state-of-the-art image synthesis approaches in both qualitative and quantitative measures (Luo et al., 2021).

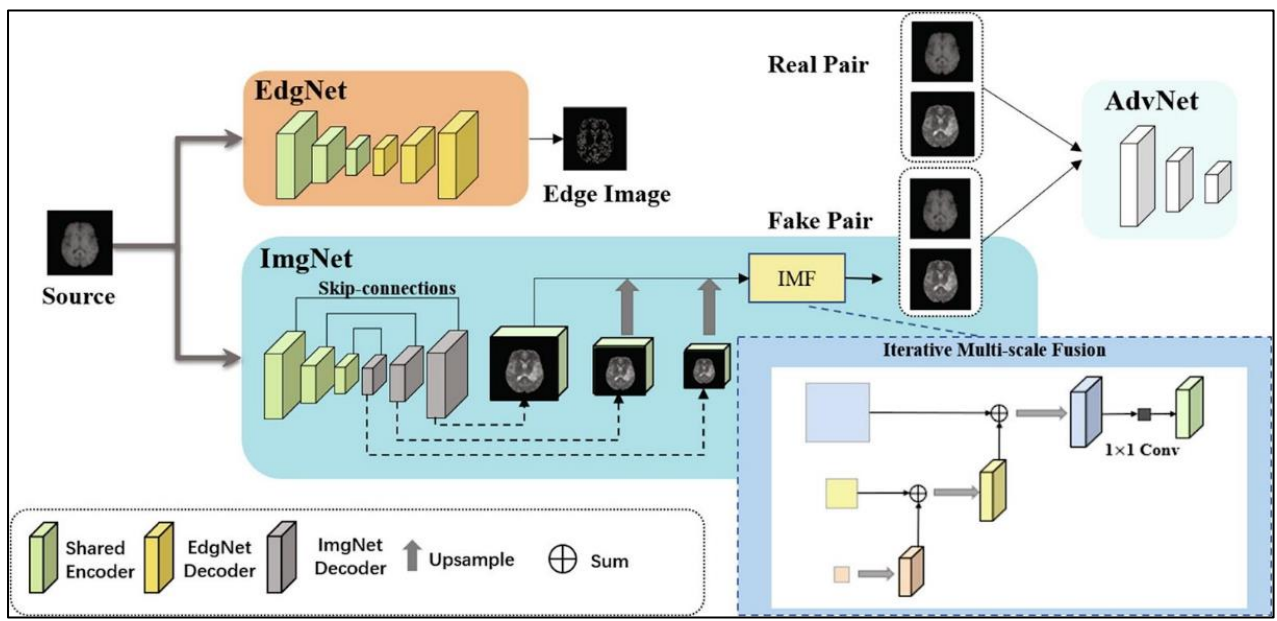

Fig 4. EP_IMG-GAN architecture (Luo et. al, 2021)

III. Brain imaging applications of Deep learning:

Deep learning has been widely used in a variety of medical imaging applications, including brain tumour detection and classification. Here's a list of the latest applications where deep learning has helped with the diagnosis and disease detection of the brain:

- Deep learning-based brain tumour classification (B Kokila et al, 2021).
- Brain tumour segmentation and grading of lower-grade glioma (Mohamed et. al, 2020).
- Classification of CT brain images to consolidate 2D images along spatial axial directions and 3D segmented blocks. (Gao et. al, 2016).
- Diagnosis of Alzheimer's disease via brain imaging (Dan et. al, 2020).
- Multimodal Brain Imaging Classification using EEG signals (Jiang et. al., 2020).
- Brain Cancer Classification (Haq et. al. ,2022).
- Cerebellum Segmentation from Foetal Brain Images (Sreelakshmy et. al, 2022).

- Semi-supervised deep learning of brain tissue segmentation. (Ito et. al, 2019).
- Developing a brain atlas. (Iqbal et. al., 2019).

IV. Future of Transformers in Medical Imaging

Transformers are the latest buzz in deep-learning research, but it is still in its nascent phase. Since the introduction of transformer architecture in 2017 in the paper by Vaswani et. al, it has become one of the most dominant deep-learning models in the research community. The following figure demonstrates transformer architecture.

Transformer models have been gaining popularity in recent years and there is a growing interest in using them in medical imaging as well. The future of Transformers in medical imaging looks promising as they have the potential to revolutionize the field by providing more accurate and efficient analysis of medical images. However, the use of Transformers in medical imaging is still in its infancy and there are several challenges that need to be addressed. They have become popular in medical imaging because of their ability to preserve contextual and edge information, which is critical for providing clinical information. Additionally, the use of self-supervised learning and multi-scale features in Transformer-based approaches has also shown promising results in cross-modality image synthesis.

The adoption of Transformers in medical imaging is limited due to the need for large amounts of well-curated training data. Obtaining labelled data for medical imaging can be expensive and difficult. The problem can be addressed by applying transfer learning aided with pretraining on a self-supervised task using large amounts of unlabelled medical data. Another challenge is the computational requirements of Transformers. These models are known to be computationally intensive and require significant computational resources. This can be a major obstacle to the widespread adoption of these models in medical imaging. Privacy concerns also need to be addressed when working with medical imaging data. Patient data is sensitive and must be always protected. This requires careful consideration of data storage, access, and sharing protocols.

The future of Transformers in medical imaging is promising and it will be interesting to see how these models continue to evolve and improve in the coming years. The use of self-supervised transfer learning with pretraining on large amounts of unlabelled medical data and the integration of edge-preserving mechanisms and multi-scale features have the potential to revolutionize medical imaging by providing a more accurate and efficient analysis of medical images.

A recent paper published in July 2022 proposes Brainformer as the latest fMRI-based hybrid Transformer Architecture for universal generalizable brain disease classification and shows good performance over different datasets and classification tasks. The results are achieved by first modelling the local cues within each brain region through 3D CNN, capturing the global relations among distant regions through two global attention blocks, and then applying a data normalization layer to handle the multisite data. (Dai et. al, 2022).

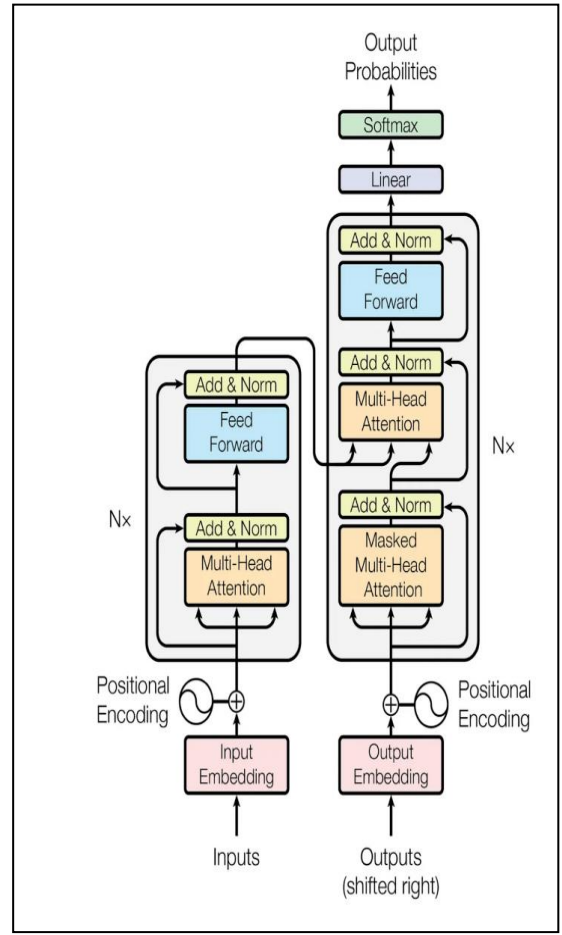

Fig 5. Transformer architecture (Vaswani et. al., 2017)

V. Future Scope and discussion

Deep learning has the potential to revolutionize medical imaging by automating the analysis of medical images and improving diagnostic accuracy. Some specific areas where deep learning is being applied or has the potential to be applied include computer-aided diagnosis, image segmentation, object, or anomaly detection, etc.

One of the most promising models for brain image synthesis can be the Transformer architecture, which is a popular choice for natural language processing tasks. The key innovation of the

Transformer is its use of self-attention mechanisms, which allows the model to weigh the importance of different parts of the input when making a prediction. This contrasts with traditional recurrent neural networks (RNNs), which rely on fixed-length context windows and can struggle to handle inputs of variable length. The Transformer's ability to efficiently process long sequences of data and its ability to be parallelized make it well-suited to tasks such as machine translation. The Transformer architecture differs from other deep learning techniques in several keyways: 1. Self-Attention, 2. Parallelization, 3. Handling sequential data, 4. Pre-training, 5. Multi-Head Attention. The Transformer architecture has been extensively used in pre-training models such as BERT and GPT-3, which have been fine-tuned for various NLP tasks with good results.

An approach that is widely getting popular is the use of Statistical shape models (SSMs) for 3D medical image segmentation. SSMs are considered a robust tool for the segmentation of medical images, and the authors reviewed the techniques around 3D SSMs. It uses landmark-based shape representations and active shape and active appearance models. They also described several alternative approaches to statistical shape modelling. However, they highlighted that establishing dense point correspondences between all shapes of the training set is generally the most challenging part of 3D model construction and one of the major factors influencing model quality, and that manual landmarking is getting increasingly unpopular due to the tedious and time-consuming expert work required and the lack of reproducibility of the results.

Another approach discussed in this review that can be useful is the use of deep learning for cross-modality MRI synthesis. The paper "Edge-preserving MRI image synthesis via the adversarial network with iterative multi-scale fusion" proposed a novel approach for cross-modality MRI synthesis using a generative adversarial network (GAN) called EP_IMF-GAN. The GAN is designed to preserve edge information, which is critical for providing clinical information, by incorporating an auxiliary task of generating the corresponding edge image of the target modality. Results from experiments on the BRATS dataset showed that the proposed EP_IMF-GAN method outperforms state-of-the-art image synthesis approaches in both qualitative and quantitative measures.

While these models, approaches, and problems have shown promising results, they also have their limitations. The Transformer architecture's reliance on the quality of the 2D images and the increased computation time and storage requirements are some of the limitations of 2D to 3D image reconstruction techniques. Additionally, SSM's reliance on manual landmarking and the lack of reproducibility of the results are some of the limitations of SSM. The limitations of TCD such as its operator dependency and the inadequate acoustic windows prevalent in certain populations can hinder its more widespread use. It is understood that deep learning-based approaches have a huge potential to impact the field of medical image analysis by improving their speed, accuracy, and efficiency and greatly improving patients' lives in the future, but it needs to go a long way to become a blind-use tool for medical imaging.

VI. Conclusion

The development of advanced imaging techniques and the advent of deep learning has led to significant advancements in the field of medical imaging. In this review, we have discussed several models, approaches, and problems related to medical imaging that have been proposed in recent literature.

The review also discussed the significance of 2D to 3D image reconstruction techniques in medical imaging. The 2D to 3D image reconstruction techniques discussed include multi-planar reformatting (MPR), Volume rendering, Maximum intensity projection (MIP), Surface rendering, Shape-from-shading (SFS), Cone-beam CT, Model-based algorithms, and Deep Learning-based methods. These techniques provide a more comprehensive and detailed view of the anatomy being imaged, allowing for more accurate diagnosis and treatment planning. The benefits of 2D to 3D image reconstruction include improved diagnostic accuracy and treatment planning, but it also has some limitations such as increased computation time, storage requirements, and reliance on the quality of the 2D images.

It also discussed the application of deep learning in brain imaging. The authors discussed that deep learning-based techniques have been used for brain tumour detection and classification, brain tumour segmentation and grading of lower-grade glioma, classification of CT brain images, detection, and diagnosis of Alzheimer's disease via brain imaging, multimodal brain imaging classification using EEG signals, brain cancer classification, classification of tumour brain images, cerebellum segmentation from foetal brain images, and developing a brain atlas. Brain imaging applications of Deep learning include Deep learning-based brain tumour detection and classification, Brain tumour segmentation and grading of lower-grade glioma, Classification of CT brain images to consolidate 2D images along spatial axial directions, and 3D segmented blocks, Detection and diagnosis of Alzheimer's disease via brain imaging, Multimodal Brain Imaging Classification using EEG signals, Brain Cancer Classification, Classifying tumour brain images,

Cerebellum Segmentation from Foetal Brain Images, Semi-supervised deep learning of brain tissue segmentation, Developing a brain atlas.

This review has highlighted the advances and recent developments in the field of brain imaging and medical imaging. The objective was to enable beginners in the field to obtain a quick overview of the progress so far and then, look at the prospects of the work.